\documentclass[sigconf,nonacm]{acmart}
\microtypesetup{expansion=false,protrusion=false}

\usepackage{booktabs}
\usepackage{graphicx}
\usepackage{tabularx}
\usepackage{hyperref}

\begin{document}

\title{Filtering Offensive Content Changes Its Visibility but Not User Behavior: Two Randomized Controlled Trials with 200,000 Users on Nextdoor}

\author{David J. Gr\"{u}ning}
\authornote{Corresponding author: gruening@stanford.edu.}
\affiliation{%
  \institution{Stanford University}
  \city{Stanford}\country{USA}}
\affiliation{%
  \institution{University of Cambridge}
  \city{Cambridge}\country{UK}}
\affiliation{%
  \institution{Max Planck Institute for Human Development}
  \city{Berlin}\country{Germany}}

\author{Matthew Katsaros}
\affiliation{%
  \institution{Yale Law School, Social Media Governance Initiative}
  \city{New Haven}\country{USA}}

\renewcommand{\shortauthors}{Gr\"{u}ning et al.}

\begin{abstract}
We investigate the effectiveness of interventions that reduce the visibility of offensive content on the local social platform Nextdoor. Content filtering---hiding or downranking offensive content that brushes against a platform's rules without clearly breaking them---is deployed across virtually every major platform, yet almost no field evidence exists on whether it changes user behavior. We report two large-scale randomized controlled trials, each involving 100,000 users. \textbf{Study 1} (2022) tested a report-triggered filter applied to comments in post threads and produced a modest 12\% reduction in views of offensive comments; across eleven further measures of platform behavior we found no significant effects. \textbf{Study 2} (2023--2024) remedied Study 1's central limitation---a weak manipulation driven by slow, report-based eligibility---by proactively scoring posts and comments at creation with Google Jigsaw's Perspective API and filtering them from the newsfeed. This produced a near-complete (95\%) reduction in views of offensive posts, yet across thirteen further measures we again found no significant effects. Across two independent trials spanning different content types, filtering mechanisms, classifiers, and countries---and despite manipulation strength rising from 12\% to 95\%---filtering reliably reduced the visibility of offensive content without altering platform visitation, content consumption, or content production. These convergent null results provide rare field evidence on a ubiquitous intervention and underscore the complexity of effectively moderating online platforms.
\end{abstract}

\ccsdesc[500]{Human-centered computing~Empirical studies in collaborative and social computing}
\ccsdesc[300]{Human-centered computing~Social media}

\keywords{content moderation, online governance, algorithmic intervention, comment filtering, harmful content, randomized controlled trial, prosocial design}

\maketitle
\textbf{Note:} This paper combines and supersedes two earlier preprints by the same authors, which reported the two studies separately: (1) Offensive comment filtering impact on online engagement - A large-scale randomized controlled trial on Nextdoor, \url{https://doi.org/10.31234/osf.io/nxuqy}, and (2) Testing the impact of filtering harmful posts: A randomized-controlled trial with 100.000 users on Nextdoor, \url{https://doi.org/10.31234/osf.io/5dy97_v1}. The present manuscript integrates both randomized controlled trials into a single paper, as Study 2 was an advance on Study 1. This paper should be cited in place of the two earlier versions. No new data were collected, the earlier preprints remain available for the record but were updated to note respectively that they are superseded by this version.

\section{Introduction}
Online platforms of all sizes spend considerable resources managing unwanted content and behavior from their users, ranging from harassment and stalking to coordinated disinformation and sexually explicit content. The most familiar response is formal content moderation: users or platform algorithms report content, which is then reviewed against a set of rules, with various remedies applied to content that violates them.

Yet a much larger volume of content brushes right up against a platform's rules without ever crossing them, and platform design tends to give this near-violating content an outsized share of engagement. As Mark Zuckerberg observed in a 2018 note, no matter where a platform draws the line for what is allowed, engagement with a piece of content rises on average as it approaches that line---even when users report afterward that they dislike it~\citep{zuckerberg2018}. Common techniques for managing such content include hiding or downranking it, adding informative labels, prompting users before content is created, and demonetizing creators.

While these interventions are ubiquitous, there is remarkably little empirical evidence on whether they work. Practitioners overwhelmingly rely on testing by other companies or within their own organization, and most public-facing evidence lives in corporate blog posts and press interviews with few experimental details~\citep{gruening2024framework}. Whether these interventions actually achieve their stated goals---and through what mechanism, and for whom---is largely unknown. Given that the ``Trust and Safety'' teams building them are typically small and responsible for a wide array of problems, it is critical that scarce resources go toward interventions that are genuinely effective.

In this paper we present evidence from two large-scale randomized controlled trials on the local social platform Nextdoor, both testing interventions that reduce the visibility of offensive content---an approach adopted by Twitter, Instagram, Facebook, YouTube, and TikTok, among others, but with little empirical grounding. Each trial involved 100,000 users and a wide range of measured outcomes, including platform visitation, content consumption, and content production.

We designed the two studies so that the second directly addresses the principal limitation of the first. Study 1 filtered comments in post threads through a report-triggered pipeline; because a comment had to be reported before it could be filtered, the intervention removed only a modest share of offensive views (a 12\% reduction), and a skeptic could reasonably attribute the downstream null results to a weak manipulation. Study 2 removed exactly this objection by scoring posts and comments proactively at creation time and filtering them from the newsfeed, achieving a near-complete (95\%) reduction in views of offensive posts. The central result is that the downstream findings did not change: in both studies, reducing the visibility of offensive content did not significantly affect site visits, content consumption, or content production, including the creation of offensive content. Reading the two trials together---with manipulation strength escalating from 12\% to 95\% and yet the same behavioral nulls---is far more informative than either study alone.

\section{Related Work}
\subsection{Detecting Offensive Content}
Media scholars and computer scientists have long studied the varying definitions, perceptions, and networked structure of offensive and toxic online content~\citep{myerswest2018, jiang2021, kumpel2023, sude2023, klein2024}. Surveys have compared perceptions of offensiveness across countries and distinguished uncivil from intolerant comments, while critical work has interrogated the ``health'' and ``toxicity'' metaphors that shape platform justifications~\citep{gibson2024}. A large strand of computer-science and HCI research focuses on detecting toxic content at scale~\citep{lees2022, villatecastillo2024}, exposing and mitigating annotation and algorithmic bias~\citep{goyal2022}, extending language coverage~\citep{buzelin2024, eskelinen2023}, and guarding against veiled speech designed to evade detection~\citep{alexiou2023, gargee2022}.

In 2017, Google's Jigsaw released the Perspective API, a free tool that returns a ``toxicity'' score for a piece of text and is now widely used by platforms and researchers alike, despite documented limitations and biases~\citep{jigsaw2025a}. In an extensive longitudinal study spanning eight platforms and 34 years, \citet{avalle2024} used Perspective to show that certain patterns in the emergence of toxic content persist across very different platform designs, pointing to the importance of underlying human behavior.

\subsection{Intervening on Offensive Content}
A parallel literature asks what can be done once offensive content is identified. Some interventions act upstream, before content is created: in a large randomized trial on Twitter, \citet{katsaros2022} found that prompting authors to reconsider offensive replies both reduced sending in the moment and lowered future offensive tweeting. Others intervene at the point of joining a community~\citep{matias2019, kim2022}; on Nextdoor, \citet{kim2022} showed that a single screen of prescriptive prosocial guidelines reduced comments reported for review. A further line seeks to reshape behavior through transparency and education following the removal of content~\citep{jhaver2019, tyler2021}.

The most prominent lever remains outright removal of rule-violating content and users~\citep{gillespie2018, roberts2019}, though its effects are hard to study because platforms release removal data only in aggregate. High-profile cases have provided natural experiments: the removal of a major political account was associated with a drop in follower toxicity and engagement (M\"{u}ller \& Schwarz, 2023); Reddit's 2015 ban of high-profile subreddits reduced hate speech among users who stayed~\citep{chandrasekharan2017}; and Germany's NetzDG regulation was associated with reduced hateful posts and hate crimes (Jim\'{e}nez-Dur\'{a}n et al., 2024).

\subsection{Reducing Content Visibility Without Removal}
Most relevant to the present work is the increasingly prevalent practice of reducing the visibility of ``borderline'' content that remains offensive without breaking a platform's rules~\citep{gillespie2018}. These actions typically occur without notice to either viewer or author, embedded in opaque algorithmic ranking, which has led users to theorize about ``shadowbanning'' even as platforms have often denied the practice~\citep{myerswest2018, lapowsky2018}. \citet{hortaribeiro2023} used a quasi-experimental design on a Facebook intervention that either filtered from view or removed comments (removal, critically, coming with an author notice) based on a continuous rule-violation score. They found that removal reduced the likelihood that an author would post similar comments in the future, whereas filtering had no significant effect on any of their measures. Their filtering null---measured on only two outcomes and given little attention in their discussion---is the finding our two studies most directly extend, across a far broader set of users and behaviors.

\section{The Present Research}
The interventions we evaluate resemble those platforms announce with hints about improved discourse and participation. When Instagram launched automatic comment filtering, its founder framed it as a response to toxic comments that discourage people from enjoying the platform and expressing themselves freely~\citep{systrom2017}. When TikTok limited comment visibility, its policy director emphasized promoting respectful discussion~\citep{wadhwa2021}. From these stated goals we derive three research questions, tested in both studies:
\begin{itemize}
  \item \textbf{RQ1.} Does filtering offensive content from view increase participation in online conversations?
  \item \textbf{RQ2.} Does filtering offensive content from view decrease the creation of similarly offensive content?
  \item \textbf{RQ3.} Does filtering offensive content from view increase visitation or general engagement with the platform?
\end{itemize}

We address these questions with two randomized controlled trials on Nextdoor, a platform on which members join with their real name and home address and interact only with others in nearby offline neighborhoods, and whose stated mission is to ``cultivate a kinder world where everyone has a neighborhood they can rely on.'' Study 1, fielded in 2022, tested a report-triggered filter on comments. Study 2, fielded in 2023--2024, tested a proactive, classifier-based filter on posts and comments in the newsfeed, and was designed as an advance on Study 1 in three respects: it scored content at creation rather than waiting on user reports, it extended filtering from comments to posts on the platform's most-used surface, and it broadened the outcome set. Both studies share an analysis strategy, described once and applied to each in turn.

\subsection{Design Overview}
Table~\ref{tab:design} summarizes how the two trials differ. The escalation in the strength and mechanism of the manipulation---from a slow, report-dependent 12\% reduction to a proactive, near-complete 95\% reduction---is the axis along which the two studies are meant to be read.

\begin{table*}[t]
  \caption{Design comparison of the two randomized controlled trials.}
  \label{tab:design}
  \small
  \begin{tabularx}{\textwidth}{lXX}
    \toprule
     & \textbf{Study 1 (2022)} & \textbf{Study 2 (2023--2024)} \\
    \midrule
    Target content & Comments in post threads & Posts and comments in the newsfeed \\
    Eligibility mechanism & Report-triggered: a member reports a comment, then Nextdoor's removal-prediction model scores it & Proactive: scored at creation via Perspective API ``Toxicity'' $\geq 0.70$ \\
    Rollout & Staggered, five waves, Aug--Oct 2022 & Single rollout from Nov 28, 2023 \\
    Population & US members & English-language members in US, Canada, UK, Australia \\
    Sample & 100,000 (50k test / 50k control) & 100,000 (50k test / 50k control) \\
    Outcome measures & 12 & 15 \\
    Observation window & 30 days pre/post (to 154 days, Appendix~\ref{app:fulltime}) & 60 days pre/post (to 280 days in supplement) \\
    Manipulation achieved & 12\% reduction in offensive-comment views & 95\% reduction in offensive-post views \\
    Downstream effects & None across 11 other measures & None across 13 other measures \\
    \bottomrule
  \end{tabularx}
\end{table*}

\subsection{Analysis Strategy (Both Studies)}
For each study we used a two-way mixed ANOVA to estimate the effect of the intervention on each outcome. One factor is condition (treatment vs.\ control) and the other is time (pre vs.\ post experiment enrollment); the interaction isolates any change in the treatment group relative to control following rollout. Because participants are enrolled upon an eligible platform visit, every participant is active on their enrollment day, producing an activity spike at day 0 that appears in both conditions; modeling time as a factor absorbs this and any other shared temporal shift, leaving the interaction to capture the intervention effect. A sensitivity analysis ($\alpha = .05$, power $\geq .95$) indicates that effects as small as $d = .02$ can be reliably detected. We corroborate each frequentist interaction test with a Bayesian ANOVA (\texttt{anovaBF} in the R package \texttt{BayesFactor}~\citep{morey2024}) using the default Jeffreys--Zellner--Siow prior, reporting $\mathit{BF}_{10}$ for the time $\times$ condition interaction; for computational tractability the Bayesian analyses use random subsamples ($N = 1{,}000$).

\section{Study 1: Report-Triggered Comment Filtering}
\subsection{Method}
\begin{figure}[t]
  \centering
  \includegraphics[width=0.7\columnwidth]{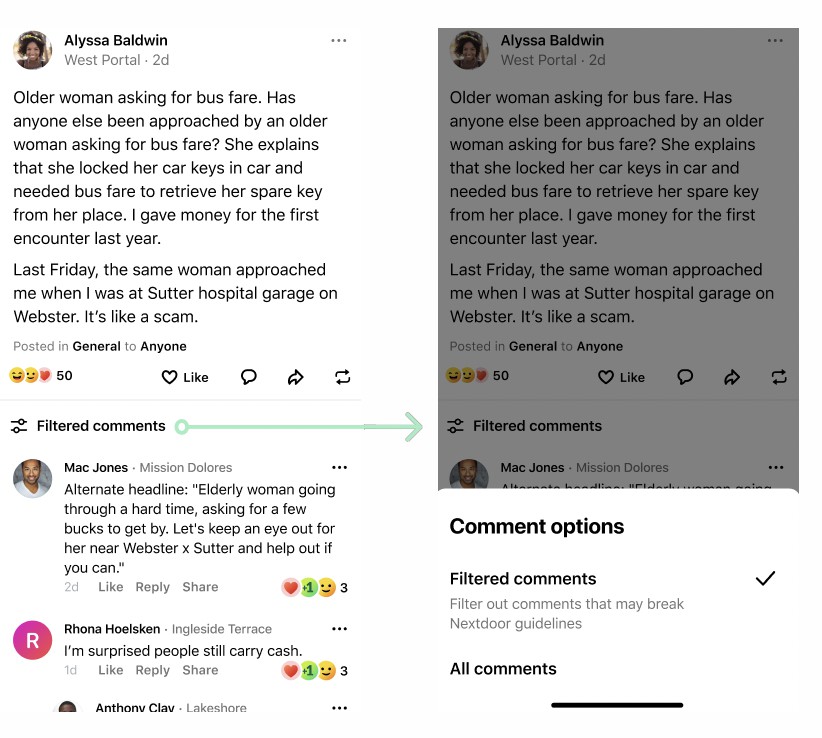}
  \caption{The comment-filter interface shown to test-condition members. A per-post toggle lets a member switch from ``Filter Comments'' (default) to ``All Comments.'' The same interface was used for filtered comments in Study 2.}
  \label{fig:interface}
\end{figure}

\textbf{Interface and eligibility.} Nextdoor built a model that detects and filters offensive comments from post threads. For members in the test group, eligible comments were filtered by default; a per-post toggle let a member switch from ``Filter Comments'' (default) to ``All Comments'' (Figure~\ref{fig:interface}), but this option was used rarely (in under 1\% of the instances it was shown). A comment was eligible for filtering if it had been reported by at least one member for a reason deemed hurtful or harmful and scored above a threshold on a Nextdoor model trained to predict whether community moderators would vote to remove it.

\textbf{Experiment design.} The feature was released through a staggered randomized rollout in five waves (beginning 2022-08-10, 08-13, 09-07, 09-30, and 10-12). A member was enrolled upon visiting a post thread that contained a filter-eligible comment, at which point they were randomized in equal proportions to test (eligible comments filtered) or control (all comments visible).

\textbf{Participants.} Participants were US Nextdoor members. From all enrolled members we sampled 10,000 from each condition in each of the five waves, for 100,000 users total.

\textbf{Measures.} We took a deliberately broad approach, measuring outcomes for all members---those who view, engage with, and create content---rather than only offensive-content authors, and across a wide set of behaviors. Twelve measures fall into four groups: a manipulation check (views of reported offensive comments); platform sessions; content consumption (comment views, post views, post-and-comment views, ad views); and content production (posts created, comments created, posts-and-comments created, offensive posts created, offensive comments created, and reactions given). Offensive posts and comments here are those reported by at least one member as hurtful or harmful; note this differs from the eligibility rule in that the manipulation-check measure is report-based.

\subsection{Results}
\textbf{Manipulation check.} As expected, we found a significant time $\times$ condition interaction on views of offensive comments (\(F(1, 99998) = 101.934\), \(p = 5.89\mathrm{e}{-}24\), \(\eta^2 = 1.54\mathrm{e}{-}4\)), confirming that the filter reduced visibility (Table~\ref{tab:s1anova}). The magnitude, however, was smaller than one might expect for an intervention meant to hide all offensive comments: a 12\% reduction relative to control. The most likely explanation is the report-triggered design---comments are not scored proactively, so much of a comment's viewing occurs before it is ever reported and filtered; in our data the 98th percentile of the time between comment creation and filtering was one week.

\textbf{Content consumption.} Across comment views, post views, post-and-comment views, and ad views we found no significant condition effect ($p$ between .169 and .794) and, critically, no significant time $\times$ condition interaction ($p$ between .160 and .556), although time main effects were significant for several measures (e.g., post-and-comment views, \(F(1, 99998) = 249.983\), \(p = 3.06\mathrm{e}{-}56\)). Bayesian analyses returned moderate-to-strong evidence for the null interaction across all four measures ($\mathit{BF}_{10}$ roughly 0.07--0.24; Table~\ref{tab:s1bayes}).

\textbf{Content production.} Across posts, comments, posts-and-comments, offensive posts, offensive comments, and reactions we again found no significant condition effect ($p$ between .239 and .876) and no significant interaction ($p$ between .379 and .935), with significant time effects on all measures except offensive posts created. Bayesian evidence again favored the null interaction throughout.

\textbf{Platform sessions.} We found no significant condition effect ($p = .132$) and no significant interaction ($p = .282$), with a significant time effect (\(F(1, 99998) = 911.549\), \(p = 2.40\mathrm{e}{-}199\)).

\textbf{Heterogeneous effects.} Binning users by pre-experiment visitation (low/med/high), content creation (none/some), and offensive-comment viewing (none/some) and rerunning the analysis within bins yielded no strong support for heterogeneous effects (Appendix~\ref{app:het1}).

\begin{table*}[t]
  \caption{Study 1: pre- and post-experiment means (SD) for all twelve measures; $n = 50{,}000$ per condition.}
  \label{tab:s1summary}
  \scriptsize
  \begin{tabularx}{\textwidth}{llXXXX}
    \toprule
    Group & Measure & Control (pre) & Test (pre) & Control (post) & Test (post) \\
    \midrule
    Manip.\ Check & Offensive Comment Views & 2.251 (6.891) & 2.294 (6.590) & 2.354 (6.359) & 2.070 (6.416) \\
    Sessions & Platform Sessions & 23.018 (36.612) & 23.434 (38.002) & 24.990 (37.879) & 25.271 (39.093) \\
    Consumption & Comment Views & 329.67 (2362.84) & 316.99 (804.17) & 359.73 (1835.90) & 364.50 (2550.62) \\
    Consumption & Post Views & 440.41 (2370.99) & 439.63 (1113.09) & 474.28 (1235.24) & 479.50 (1216.10) \\
    Consumption & Post + Comment Views & 197.96 (455.01) & 201.01 (458.57) & 211.30 (496.33) & 212.17 (473.15) \\
    Consumption & Ad Views & 58.35 (140.15) & 59.79 (145.22) & 61.70 (145.98) & 62.70 (146.50) \\
    Production & Posts Created & 0.132 (1.076) & 0.128 (0.870) & 0.118 (1.133) & 0.114 (0.927) \\
    Production & Comments Created & 0.652 (5.252) & 0.635 (5.060) & 0.697 (5.236) & 0.692 (5.487) \\
    Production & Posts + Comments Created & 0.784 (5.821) & 0.763 (5.496) & 0.815 (5.870) & 0.806 (5.895) \\
    Production & Offensive Posts Created & 0.002 (0.062) & 0.001 (0.057) & 0.002 (0.062) & 0.001 (0.050) \\
    Production & Offensive Comments Created & 0.009 (0.200) & 0.008 (0.178) & 0.012 (0.403) & 0.013 (0.271) \\
    Production & Reactions Given & 1.833 (13.455) & 1.748 (13.423) & 1.998 (13.659) & 1.955 (14.496) \\
    \bottomrule
  \end{tabularx}
\end{table*}

\begin{table*}[t]
  \caption{Study 1: two-way mixed ANOVA (condition $\times$ time). $F$ and $p$ shown for each factor; $^{*}\,p < .05$. Effect sizes ($\eta^2$) are given in Table~\ref{tab:ges1}.}
  \label{tab:s1anova}
  \scriptsize
  \begin{tabular}{llrrrrrr}
    \toprule
     & & \multicolumn{2}{c}{Condition} & \multicolumn{2}{c}{Time} & \multicolumn{2}{c}{Interaction} \\
    \cmidrule(lr){3-4}\cmidrule(lr){5-6}\cmidrule(lr){7-8}
    Group & Measure & $F$ & $p$ & $F$ & $p$ & $F$ & $p$ \\
    \midrule
    Manip. & Offensive Comment Views & 9.919 & .002$^{*}$ & 13.917 & 1.9e-04$^{*}$ & 101.934 & 5.9e-24$^{*}$ \\
    Sessions & Platform Sessions & 2.274 & .132 & 911.549 & 2.4e-199$^{*}$ & 1.156 & .282 \\
    Consump. & Comment Views & 0.162 & .687 & 23.253 & 1.4e-06$^{*}$ & 1.175 & .278 \\
    Consump. & Post Views & 0.068 & .794 & 52.477 & 4.4e-13$^{*}$ & 0.347 & .556 \\
    Consump. & Post + Comment Views & 0.466 & .495 & 249.983 & 3.1e-56$^{*}$ & 1.974 & .160 \\
    Consump. & Ad Views & 1.894 & .169 & 173.373 & 1.5e-39$^{*}$ & 0.853 & .356 \\
    Prod. & Posts Created & 0.492 & .483 & 33.112 & 8.7e-09$^{*}$ & 0.007 & .933 \\
    Prod. & Comments Created & 0.126 & .723 & 16.710 & 4.4e-05$^{*}$ & 0.233 & .629 \\
    Prod. & Posts + Comments Created & 0.196 & .658 & 7.433 & .006$^{*}$ & 0.187 & .666 \\
    Prod. & Offensive Posts Created & 1.385 & .239 & 0.966 & .326 & 0.007 & .935 \\
    Prod. & Offensive Comments Created & 0.024 & .876 & 13.468 & 2.4e-04$^{*}$ & 0.158 & .691 \\
    Prod. & Reactions Given & 0.592 & .442 & 61.116 & 5.4e-15$^{*}$ & 0.774 & .379 \\
    \bottomrule
  \end{tabular}
\end{table*}

\begin{table}[t]
  \caption{Study 1: Bayes factors ($\mathit{BF}_{10}$) for the time $\times$ condition interaction across five random subsamples ($N = 1{,}000$). Values well below 1 indicate moderate-to-strong evidence for the null.}
  \label{tab:s1bayes}
  \small
  \begin{tabular}{llr}
    \toprule
    Group & Measure & $\mathit{BF}_{10}$ (min--max) \\
    \midrule
    Sessions & Platform Sessions & 0.070--0.269 \\
    Consumption & Comment Views & 0.075--0.124 \\
    Consumption & Post Views & 0.072--0.137 \\
    Consumption & Post + Comment Views & 0.084--0.136 \\
    Consumption & Ad Views & 0.073--0.242 \\
    Production & Posts Created & 0.070--0.111 \\
    Production & Comments Created & 0.071--0.133 \\
    Production & Posts + Comments Created & 0.070--0.115 \\
    Production & Offensive Posts Created & 0.069--0.238 \\
    Production & Offensive Comments Created & 0.071--0.205 \\
    Production & Reactions Given & 0.070--0.171 \\
    \bottomrule
  \end{tabular}
\end{table}

\begin{figure*}[t]
  \centering
  \includegraphics[width=\textwidth]{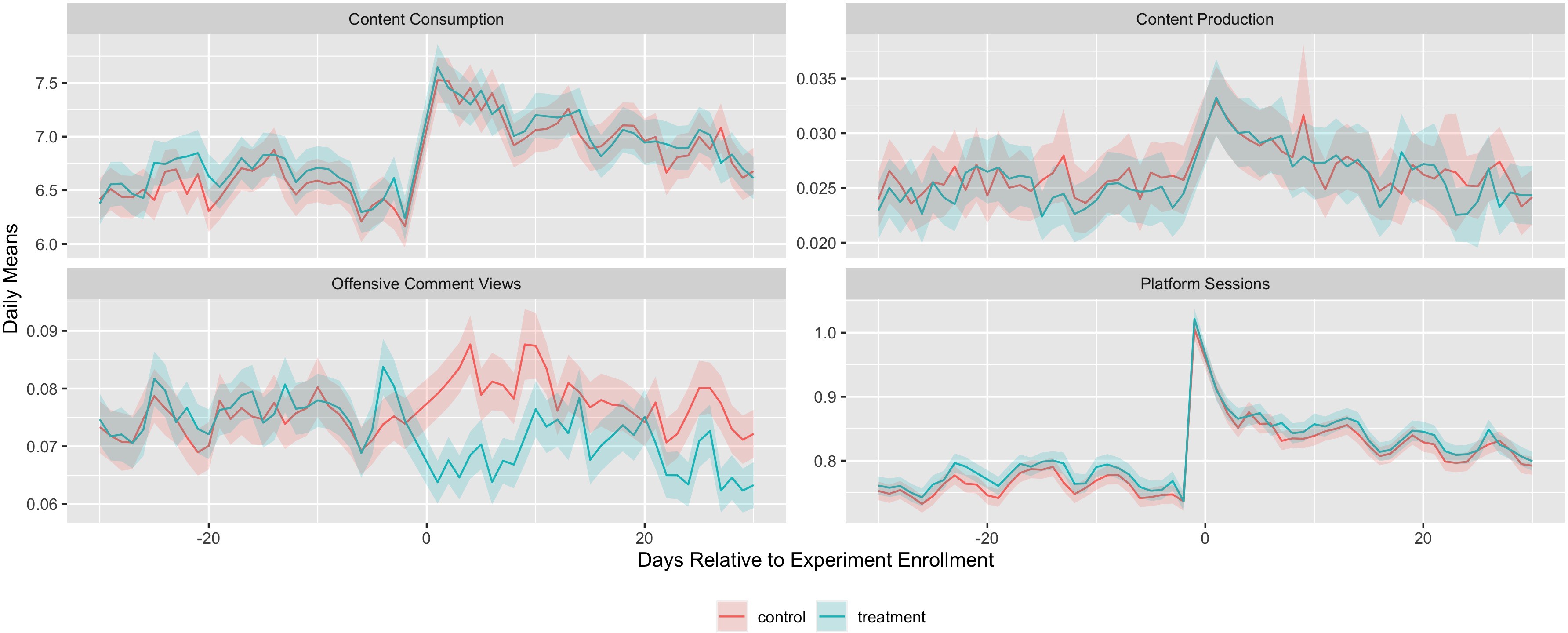}
  \caption{Study 1: daily means ($\pm$95\% CI) for platform sessions, posts and comments created, posts and comments viewed, and offensive-comment views, 30 days before and after enrollment. Control in red, treatment in teal. Only offensive-comment views differs significantly between conditions.}
  \label{fig:s1results}
\end{figure*}

\subsection{Interim Discussion: A Weak Manipulation Motivates Study 2}
Study 1 shows that report-triggered filtering reduces the visibility of offensive comments but does not detectably change any other platform behavior. The natural objection to this null is that the manipulation was weak: a 12\% reduction leaves most offensive views intact, so perhaps a stronger intervention would move downstream outcomes. Study 2 was designed to test exactly this. If the behavioral nulls are an artifact of a weak manipulation, a near-complete reduction should reveal the effects Study 1 missed; if instead the nulls reflect something more fundamental about opaque, visibility-only moderation, they should persist even when almost all offensive views are eliminated.

\section{Study 2: Proactive Newsfeed Filtering of Posts and Comments}
\subsection{Method}
\textbf{Filter eligibility.} Study 2 filtered offensive posts and comments from view in the newsfeed. Almost immediately after creation, content was scored with Google Jigsaw's Perspective API ``Toxicity'' attribute---a rude, disrespectful, or unreasonable comment that is likely to make people leave a discussion~\citep{jigsaw2025b}---which returns a score between 0 and 1~\citep{jigsaw2025c}. Content scoring at or above 0.70 was eligible for filtering and hidden from the newsfeed for the test condition; content below threshold was unaffected. Only content by members with an English language setting was scored. This proactive, classifier-based rule replaces Study 1's dependence on member reports and thus targets offensive content before it accrues views; prior work suggests even lower thresholds can be effective~\citep{avalle2024}.

\textbf{Experiment design.} Beginning November 28, 2023, members were enrolled upon visiting any part of Nextdoor and simultaneously randomized in equal proportions to test or control. Test-condition members had eligible posts and comments filtered from the newsfeed; on-platform and email notifications that would have surfaced filter-eligible posts were suppressed; and filter-eligible comments were hidden by default when expanding a post's comments. Control members saw all content and received all notifications.

\textbf{Participants.} Because scoring applied only to English content, members in the US, Canada, UK, and Australia were eligible; enrollment additionally required a visit between November 28 and December 2, 2023. We sampled 50,000 test and 50,000 control members (100,000 total).

\textbf{Measures.} Following the same broad philosophy as Study 1, we collected fifteen measures per participant per day across four groups: a manipulation check (views of offensive posts and, separately, offensive comments); platform sessions; content consumption (posts viewed, comments viewed, ads viewed, viewed posts the participant reported, and the average Perspective toxicity of posts and of comments viewed); and content production (posts created, comments created, reactions given, offensive posts created, offensive comments created, and posts/comments created that another member reported). Individual reaction types were consistent with their sum, so we report the sum. Offensive created content uses a lower threshold (Toxicity $\geq 0.60$) than offensive viewed content. We analyze 60 days pre- and post-enrollment (120 days per participant); a supplement extends this to 80/200 days on a 40,000-participant subsample with consistent results.

\subsection{Results}
\textbf{Manipulation check.} The proactive filter produced a strong manipulation. For offensive-post views we found a significant time $\times$ condition interaction (\(F(1, 99998) = 635.725\), \(p = 7.80\mathrm{e}{-}140\), \(\eta^2 = 0.002\)), corresponding to a near-complete (95\%) reduction relative to control. For offensive-comment views we likewise found a significant interaction (\(F(1, 99998) = 603.446\), \(p = 7.40\mathrm{e}{-}133\), \(\eta^2 = 0.001\)). This is the key contrast with Study 1: the intervention now removes almost all offensive views rather than 12\% (Table~\ref{tab:s2anova}).

\textbf{Content consumption.} Across posts viewed, comments viewed, ads viewed, viewed posts reported, and the average toxicity of posts and comments viewed, we found no significant time $\times$ condition interaction for five of the six measures ($p$ between .081 and .954). The single exception was the average toxicity of comments viewed (\(F(1, 99998) = 7.779\), \(p = .005\), \(\eta^2 = 4.07\mathrm{e}{-}6\))---statistically significant but modest (a 4.8\% pre-to-post decrease in the test condition vs.\ 2.6\% in control). Bayesian analyses returned moderate-to-strong evidence for the null interaction across measures (Table~\ref{tab:s2bayes}). In short, filtering changed the volume of consumption not at all, and the quality of consumption only slightly.

\textbf{Content production.} Across posts created, comments created, reactions given, offensive posts created, offensive comments created, and reported posts/comments, we found no significant interaction for any measure ($p$ between .243 and .985). Notably, this includes no reduction in the creation of offensive content---the outcome platforms often hope such interventions will shift.

\textbf{Platform sessions.} We found no significant condition effect ($p = .478$) and no significant interaction ($p = .45$), with a small significant time effect (\(F(1, 99998) = 5.854\), \(p = .016\)).

\textbf{Heterogeneous effects.} We binned users along the same three pre-experiment dimensions as Study 1 and reran the analysis within seven cohorts. Results were highly consistent with the main analysis: the manipulation check was always significantly reduced, the average toxicity of comments viewed was slightly reduced in four of seven cohorts, and only one cohort (very active visitors) showed an additional small significant effect. No cohort showed changes in platform use, consumption, or production (Appendix~\ref{app:het2}).

\begin{table*}[t]
  \caption{Study 2: pre- and post-experiment means (SD) for all fifteen measures; $n = 50{,}000$ per condition. (Carried over from the second manuscript.)}
  \label{tab:s2summary}
  \centering
  \includegraphics[width=\textwidth]{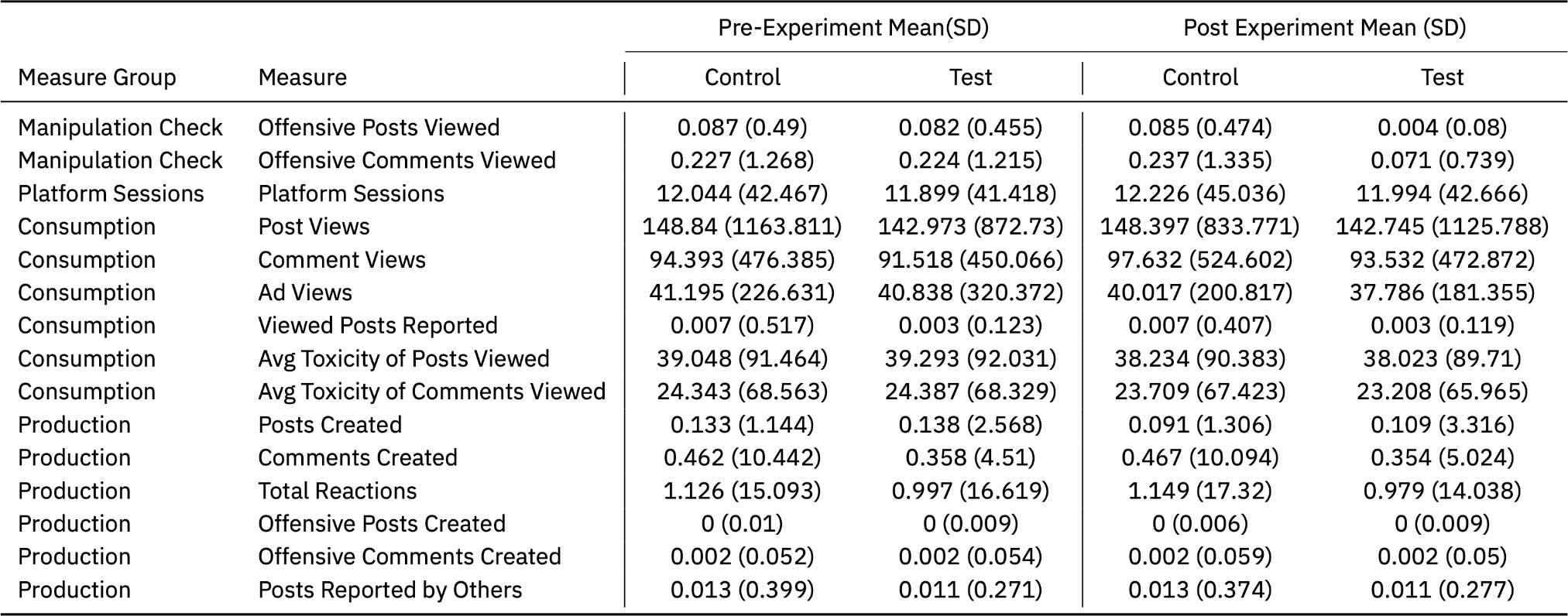}
\end{table*}

\begin{table*}[t]
  \caption{Study 2: two-way mixed ANOVA (condition $\times$ time) results for all measures; $^{*}\,p < .05$. Effect sizes are given in Table~\ref{tab:ges2}.}
  \label{tab:s2anova}
  \centering
  \includegraphics[width=\textwidth]{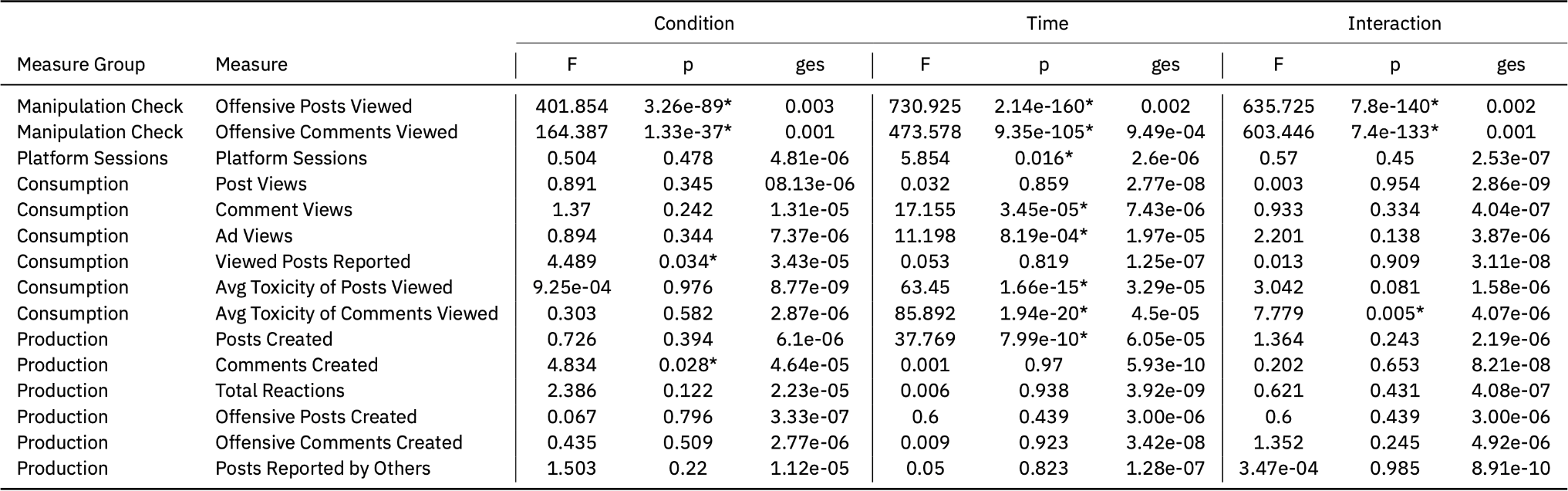}
\end{table*}

\begin{table}[t]
  \caption{Study 2: Bayes factors ($\mathit{BF}_{10}$) for the time $\times$ condition interaction ($N = 1{,}000$). The Bayes factor for viewed posts reported was not computed; see Appendix~\ref{app:bf}.}
  \label{tab:s2bayes}
  \centering
  \includegraphics[width=\columnwidth]{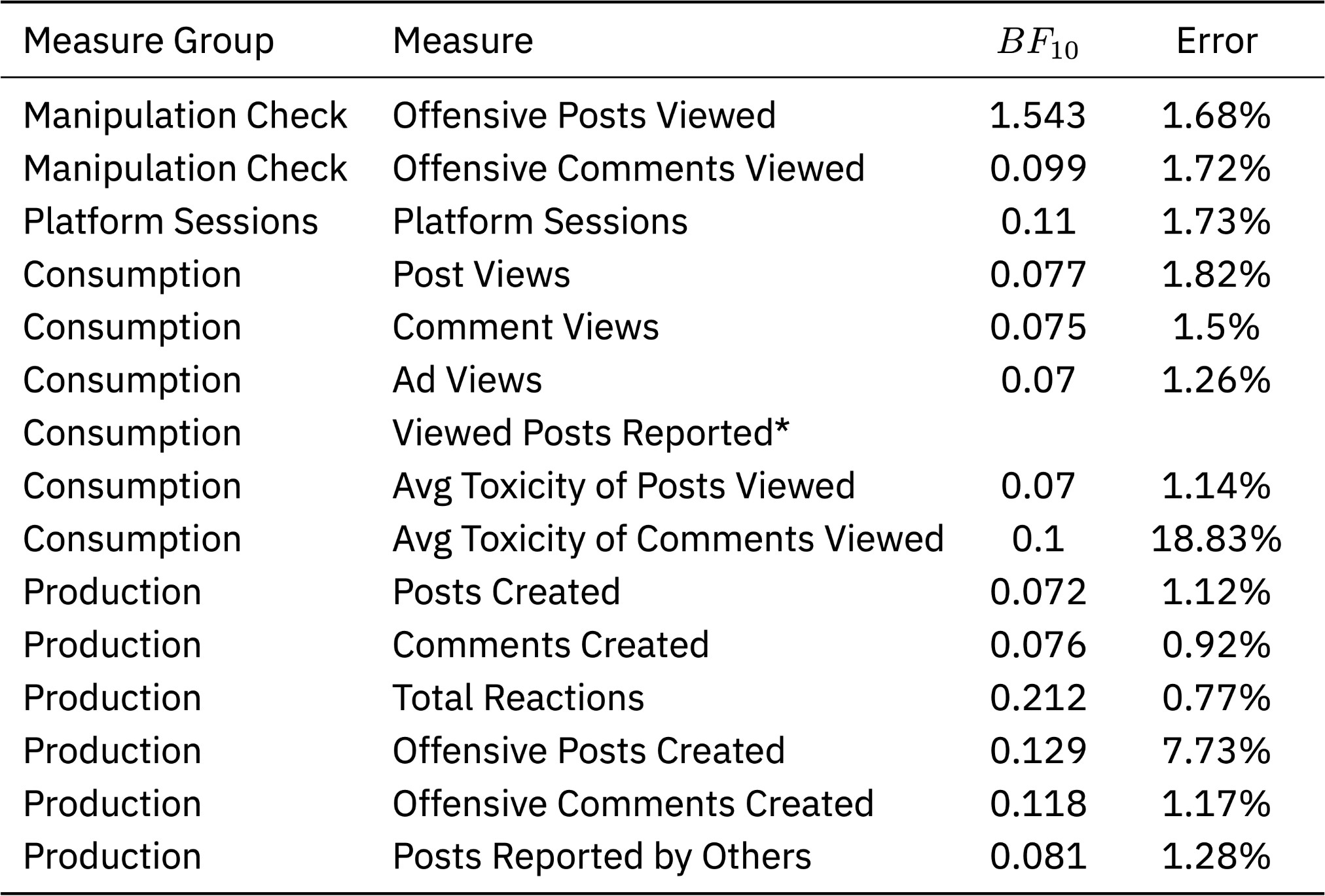}
\end{table}

\begin{figure*}[t]
  \centering
  \includegraphics[width=0.8\textwidth]{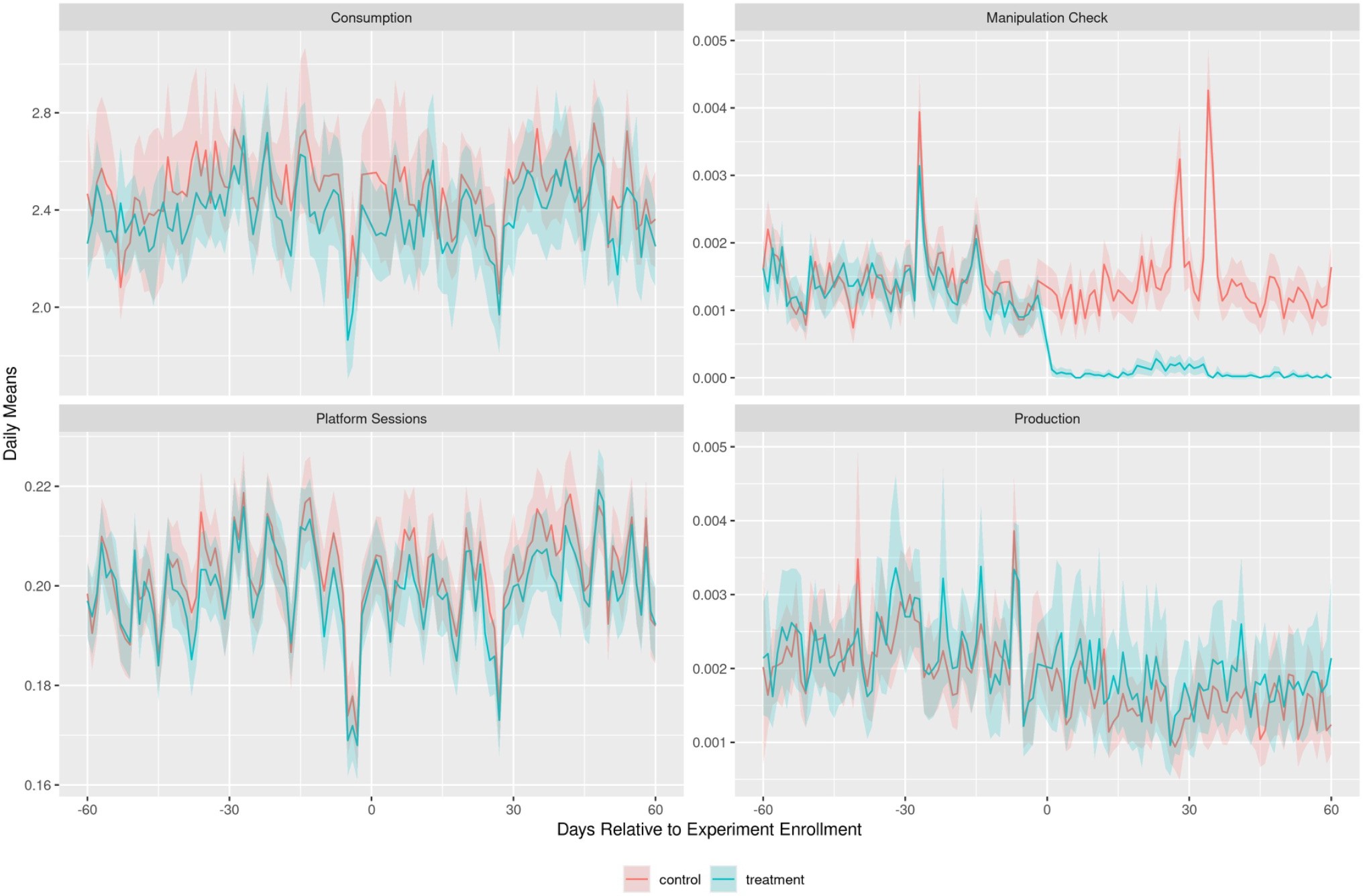}
  \caption{Study 2: daily means ($\pm$95\% CI) for platform sessions, posts created, posts viewed, and offensive-post views, 60 days before and after enrollment. Control in red, treatment in teal. Only offensive-post views differs significantly between conditions.}
  \label{fig:s2results}
\end{figure*}

\section{General Discussion}
Across two independent randomized controlled trials---different content types (comments vs.\ posts and comments), different filtering mechanisms (report-triggered vs.\ proactive), different classifiers (a proprietary removal-prediction model vs.\ the public Perspective API), different platform surfaces (post threads vs.\ the newsfeed), and different countries---we find the same result. Filtering reliably reduces the visibility of offensive content, but it does not change platform visitation, the volume of content consumption or production, or the creation of offensive content. Crucially, this holds even as the strength of the manipulation rises from a 12\% reduction in Study 1 to a 95\% reduction in Study 2. The most obvious explanation for Study 1's behavioral nulls---that the filter simply did not remove enough offensive content---is ruled out by Study 2. The nulls appear to be a property of opaque, visibility-only moderation, not of an underpowered manipulation.

These results extend \citet{hortaribeiro2023}, who found on Facebook that comment removal (which comes with an author notice) reduced future offensive comments, whereas filtering (which does not) had no effect on their two measures. We replicate their filtering null across a far broader population---including users who merely view content---and a much larger set of behaviors. Their discussion offered a technical account of the filtering-vs-removal difference but omitted the most straightforward one: removal is accompanied by user-facing messaging, whereas filtering is opaque and users are unaware the platform has acted. Our findings underscore that what matters is not only the intervention mechanism but the information users receive about it---consistent with evidence that informative interventions are essential in ``wicked'' environments where outcomes are uncertain and feedback is delayed~\citep{gruening2022wicked}, and with related work showing that transparency~\citep{jhaver2019}, education~\citep{pennycook2022}, and creation-time prompts~\citep{katsaros2022} can reshape norms and reduce offensive-content creation, aligning with emerging principles of prosocial platform design~\citep{grueningkamin2025}.

At the same time, the stability of engagement under filtering can be read as a positive null. A central worry for platforms deploying such interventions is that they will suppress engagement. Our two trials show that even pervasive filtering does not reduce user engagement across a wide range of metrics---from basic visitation to consumption to production. Filtering may therefore be a low-risk way to reduce the visibility of offensive content, even if it is not, on its own, a lever for changing behavior. This ecologically valid, on-platform test complements independent evaluation methods~\citep{gruening2025independent}.

The path forward likely combines visibility reduction with feedback to authors. Filtering offensive content while informing its creator would be expected to reduce the overall volume of offensive content, and greater transparency about platform action is increasingly required by regulation such as the EU's Digital Services Act, which obliges large platforms to notify authors when their content is removed or demoted.

These findings also bear on public accountability. When platforms are scrutinized, spokespeople frequently cite the existence of ``safety'' features as a defense---for example, Meta's response to whistleblower testimony that it had introduced over 30 tools to help teens and families have safe experiences online~\citep{ortutay2023, nix2023}. Without transparency about how such features are implemented or experimental evidence of their efficacy, it is impossible to know whether these tools help. Collaborative field studies like ours offer a rare means of finding out, complementing independent methods such as third-party apps~\citep{gruening2023onesec, haliburton2024} and browser extensions~\citep{lukoff2023}.

\subsection{Limitations}
Our measures capture platform behavior, not attitudes; filtering might shift users' perceptions of their neighbors or their well-being in ways we did not assess, which future work could capture through experience sampling. Most measures also speak to the volume rather than the quality of interactions---though our reaction-type and (in Study 2) toxicity-of-content-viewed measures found little movement, and Perspective's newer ``bridging'' scores~\citep{jigsaw2024} were available for too little of the observation window to use. Our heterogeneous-effects analyses, while consistent, are not exhaustive; dimensions such as user tenure might yet reveal differences, for instance if filtering matters more for newer members not yet inured to offensive content. Finally, both interventions were opaque by design; the transparency-augmented variants discussed above remain to be tested directly.

\subsection{Conclusion}
Many platforms use interventions like the ones tested here to reduce views of offensive content. Across two large-scale randomized trials, we found that filtering had its intended effect on visibility---modestly in Study 1 and nearly completely in Study 2---yet in neither study did it move a broad set of engagement, consumption, or production outcomes, including the creation of offensive content. Contrary to conventional wisdom, opaque visibility-reduction interventions may be effective at hiding specific content but not at altering underlying behavior or shaping the norms these interventions ultimately aim for. Above all, the effective moderation of online platforms is complex, and nuance is required in developing and, especially, evaluating digital interventions.

\begin{acks}
We thank Rafael Jim\'{e}nez-Dur\'{a}n and the Justice Collaboratory---including Tom Tyler, Tracey Meares, and Caroline Nobo---for their support and feedback, and our collaborators at Nextdoor, including Fay Johnson, Kyle Miller, Laura Bisesto, Matthew Boggess, Mark Jaskolski, Vishakha Nangia, and Sugin Lou.
\end{acks}

\section*{Data and Ethics}
Data and analysis code for both studies are available on OSF (Study 1: \url{https://osf.io/3wgqr/}; Study 2: \url{https://osf.io/qfbv6/}). Both studies were approved by the Yale University IRB (Study 1: May 30, 2023, \#2000035342; Study 2: October 8, 2024, \#2000038937).

\medskip
\noindent\textbf{Conflicts of interest.} This research was conducted in collaboration with Nextdoor, the social platform on which the interventions evaluated here were deployed. Nextdoor designed and built the content-filtering features under study and provided the de-identified platform data used in both experiments. The academic authors (D.J.G.\ and M.K.) received no personal compensation or equity from Nextdoor for this work.

\noindent\textbf{Researcher independence.} The academic authors retained full control over the study analysis, result interpretation, and the decision to publish; publication was not contingent on the direction of the results. Nextdoor reviewed the manuscript before submission only to verify the factual accuracy of the platform description and to confirm that no confidential or user-identifying information was disclosed.

\noindent\textbf{Funding.} DG’s work is funded by the Huo Family Foundation and Stanford’s Center for Digital Health.

\nocite{*}
\bibliographystyle{ACM-Reference-Format}
\bibliography{references}

\appendix

\section{Full-Timeframe Analysis (Study 1)}
\label{app:fulltime}
\begin{figure*}[t]
  \centering
  \includegraphics[width=0.9\textwidth]{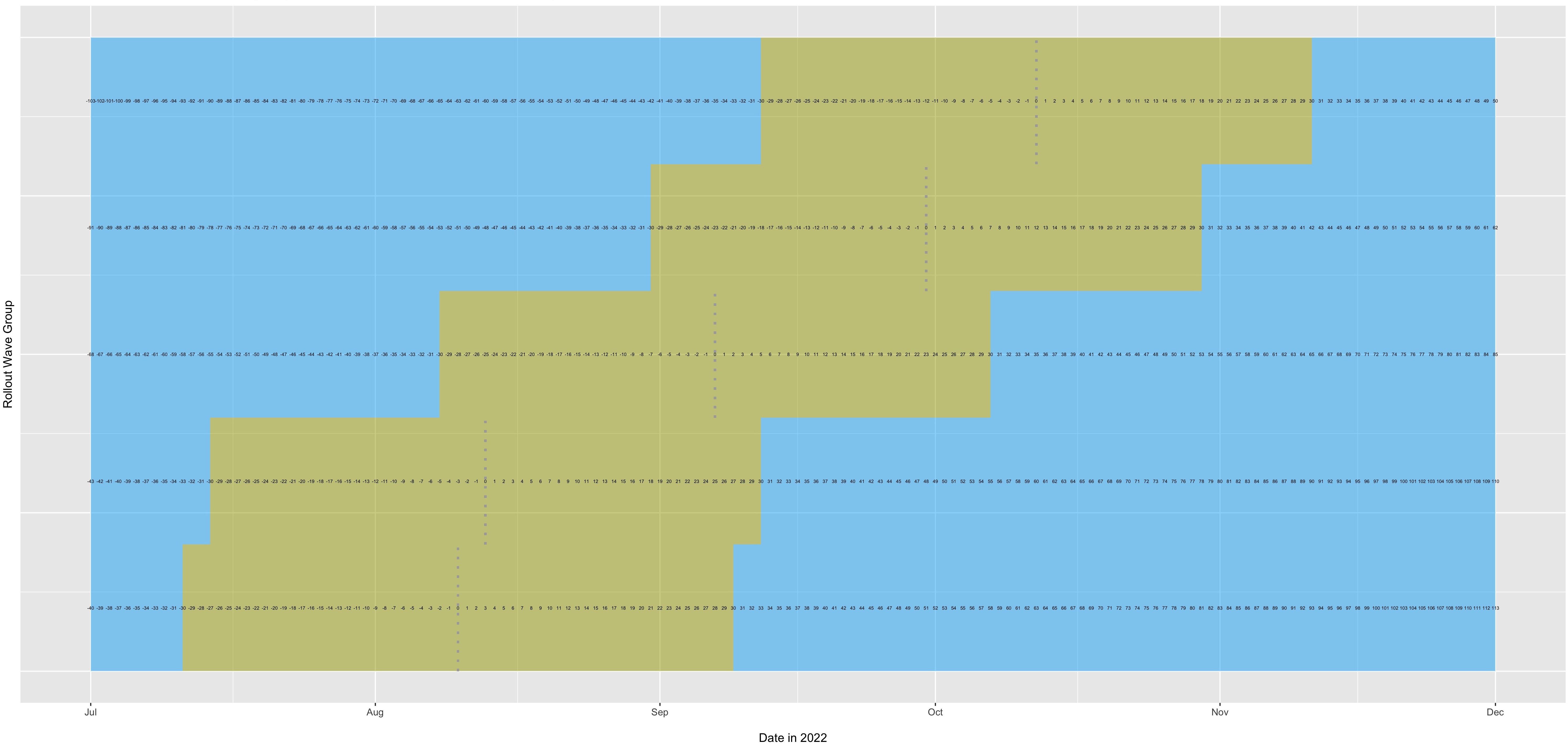}
  \caption{Coverage of the five rollout waves of Study 1 over the period for which data were available. Each row is one rollout wave, with the horizontal axis giving calendar date in 2022. Green shading marks the 30-days-before/30-days-after window analysed in the main text; the full extent (blue and green combined) is the data used in Table~\ref{tab:fulltime}.}
  \label{fig:waves}
\end{figure*}

Study 1 was rolled out in five waves, the first beginning 2022-08-10 and the last 2022-10-12. The dataset contained one row per participant per day between 2022-07-01 and 2022-11-30, regardless of the wave a participant belonged to. Consequently, participants in the earlier waves contribute relatively less pre-experiment data and relatively more post-experiment data; for every wave, at least 30 days before and after enrollment are available (Figure~\ref{fig:waves}).

The main text restricts the analysis to the 30 days before and after enrollment. Here we repeat the identical two-way mixed ANOVA using the full set of available dates, which extends to as many as 103 days before and 113 days after enrollment depending on wave.

The findings are unchanged (Table~\ref{tab:fulltime}). Every measure shows a significant time effect, indicating that platform behavior shifted over the observation period regardless of condition. As in the 30-day analysis, the only measure with a significant time $\times$ condition interaction is the manipulation check, views of offensive comments.

\begin{table*}[t]
  \caption{Study 1: two-way mixed ANOVA (condition $\times$ time) using the full set of available dates rather than the 30-day window used in the main text. $^{*}\,p < .05$. Four time-factor $p$ values were returned as exactly 0 by the analysis software (floating-point underflow) and are reported here as ${<}\,1\mathrm{e}{-}300$.}
  \label{tab:fulltime}
  \small
  \begin{tabular}{llrrrrrr}
    \toprule
     & & \multicolumn{2}{c}{Condition} & \multicolumn{2}{c}{Time} & \multicolumn{2}{c}{Interaction} \\
    \cmidrule(lr){3-4}\cmidrule(lr){5-6}\cmidrule(lr){7-8}
    Group & Measure & $F$ & $p$ & $F$ & $p$ & $F$ & $p$ \\
    \midrule
    Manip. & Offensive Comment Views & 3.041 & .081 & 317.291 & 7.3e-71$^{*}$ & 18.510 & 1.7e-05$^{*}$ \\
    Sessions & Platform Sessions & 0.974 & .324 & 6980.727 & ${<}$1e-300$^{*}$ & 0.110 & .740 \\
    Consump. & Comment Views & 0.023 & .880 & 820.970 & 8.0e-180$^{*}$ & 0.640 & .424 \\
    Consump. & Post Views & 0.121 & .728 & 1497.461 & ${<}$1e-300$^{*}$ & 1.909 & .167 \\
    Consump. & Post + Comment Views & 0.253 & .615 & 2811.672 & ${<}$1e-300$^{*}$ & 0.006 & .936 \\
    Consump. & Ad Views & 1.331 & .249 & 2679.546 & ${<}$1e-300$^{*}$ & 0.040 & .842 \\
    Prod. & Posts Created & 0.483 & .487 & 14.321 & 1.5e-04$^{*}$ & 0.557 & .455 \\
    Prod. & Comments Created & 0.030 & .861 & 242.572 & 1.3e-54$^{*}$ & 1.225 & .268 \\
    Prod. & Posts + Comments Created & 0.096 & .757 & 217.634 & 3.3e-49$^{*}$ & 1.413 & .235 \\
    Prod. & Offensive Posts Created & 1.942 & .163 & 11.797 & 5.9e-04$^{*}$ & 0.234 & .629 \\
    Prod. & Offensive Comments Created & 0.023 & .880 & 44.545 & 2.5e-11$^{*}$ & 0.059 & .808 \\
    Prod. & Reactions Given & 0.848 & .357 & 341.639 & 3.8e-76$^{*}$ & 0.163 & .686 \\
    \bottomrule
  \end{tabular}
\end{table*}

\section{Heterogeneous Effects (Study 1)}
\label{app:het1}
The main analysis compares treatment and control participants as whole groups. It is possible that the intervention helped some subgroups and hindered others, producing an aggregate null. To test this, we binned participants on their behavior in the 30 days before enrollment along three dimensions.

\textbf{Platform visitation.} Participants in the Low bin visited on 10 or fewer days ($n = 56{,}514$), the Med bin on 11--20 days ($n = 20{,}822$), and the High bin on 21--30 days ($n = 22{,}664$).

\textbf{Content production.} Participants who created no posts or comments in the 30 days before the experiment form the ``No Content Produced'' bin ($n = 81{,}769$); those who created at least one form the ``Some Content Produced'' bin ($n = 18{,}231$).

\textbf{Offensive comment views.} Participants with no views of offensive comments in the 30 days before the experiment form the ``No Offensive Comment Views'' bin ($n = 58{,}866$); those with one or more form the ``Some Offensive Comment Views'' bin ($n = 41{,}134$).

For each dimension we extended the main analysis to a three-way mixed ANOVA, adding the binning variable as a further between-subject factor. Across all three binning variables, no measure showed a significant three-way interaction ($p < .05$). As in the main analysis, the offensive-comment-views measure showed a significant time $\times$ condition interaction throughout. We therefore find no strong support for heterogeneous effects: the main findings hold across these subgroups. Code and full model output are in the ``Heterogeneous Effects Analysis'' folder of the Study 1 OSF repository (\url{https://osf.io/3wgqr/}).

\section{Heterogeneous Effects (Study 2)}
\label{app:het2}
We repeated the same exercise for Study 2, binning participants on their behavior in the 30 days before enrollment along three dimensions.

\textbf{Platform visitation.} Participants in the No bin did not visit the platform at all in the 30 days before the experiment ($N = 27{,}542$), the Med bin visited between 1 and 9 times ($N = 6{,}980$), and the High bin visited 10 or more times ($N = 5{,}478$).

\textbf{Content production.} Participants who created no posts or comments form the ``No Content Produced'' bin ($N = 37{,}311$); those who created at least one form the ``Some Content Produced'' bin ($N = 2{,}689$).

\textbf{Offensive content views.} Participants with no views of offensive posts or comments form the ``No Offensive Content Views'' bin ($N = 37{,}172$); those with one or more form the ``Some Offensive Content Views'' bin ($N = 2{,}828$).

As expected given randomization, participants were distributed consistently across these bins in the test and control conditions. Rather than adding the bin as a factor, here we repeated the two-way mixed ANOVA separately within each bin, giving seven separate cohort analyses.

Results were highly consistent with the main analysis. The manipulation-check measures---views of offensive posts and of offensive comments---were significantly reduced in every cohort. In four of the seven analyses the average Perspective toxicity of comments viewed was significantly but only slightly reduced, mirroring the main analysis. In one analysis only---participants who visited the platform 10 or more times in the 30 days before the experiment---the average toxicity of posts viewed showed an additional small significant decrease. No cohort showed a change in platform use, content consumption, or content production. Code and full model output are in the ``Heterogeneous Effects Analysis'' folder of the Study 2 OSF repository (\url{https://osf.io/qfbv6/}).

\section{Uncomputed Bayes Factor (Study 2)}
\label{app:bf}
Table~\ref{tab:s2bayes} reports Bayes factors for the time $\times$ condition interaction for each Study 2 measure. No Bayes factor is reported for one consumption measure, viewed posts reported by the participant. This measure is extremely sparse (means of 0.003--0.007 events per participant per day; Table~\ref{tab:s2summary}), and the Bayesian ANOVA was not estimated for it. The corresponding frequentist test is reported in Table~\ref{tab:s2anova} and shows no significant interaction ($F(1, 99998) = 0.013$, $p = .909$).

\section{Effect Sizes}
\label{app:effectsizes}
Tables~\ref{tab:s1anova} and~\ref{tab:s2anova} in the main text omit effect sizes for space. Generalized eta-squared values for both studies are given in Tables~\ref{tab:ges1} and~\ref{tab:ges2}.

\begin{table*}[t]
  \caption{Study 1: generalized eta-squared effect sizes for the two-way mixed ANOVA reported in Table~\ref{tab:s1anova} (30-day window).}
  \label{tab:ges1}
  \small
  \begin{tabular}{llrrr}
    \toprule
    Group & Measure & Cond. & Time & Inter. \\
    \midrule
    Manip. & Offensive Comment Views & 8.40e-05 & 2.13e-05 & 1.56e-04 \\
    Sessions & Platform Sessions & 2.12e-05 & 6.31e-04 & 8.00e-07 \\
    Consump. & Comment Views & 9.70e-07 & 9.34e-05 & 4.72e-06 \\
    Consump. & Post Views & 5.01e-07 & 1.38e-04 & 9.12e-07 \\
    Consump. & Post + Comment Views & 4.34e-06 & 1.69e-04 & 1.34e-06 \\
    Consump. & Ad Views & 1.77e-05 & 1.17e-04 & 5.77e-07 \\
    Prod. & Posts Created & 4.16e-06 & 5.08e-05 & 1.09e-08 \\
    Prod. & Comments Created & 1.08e-06 & 2.36e-05 & 3.28e-07 \\
    Prod. & Posts + Comments Created & 1.70e-06 & 1.01e-05 & 2.54e-07 \\
    Prod. & Offensive Posts Created & 7.67e-06 & 4.31e-06 & 3.00e-08 \\
    Prod. & Offensive Comments Created & 1.43e-07 & 5.59e-05 & 6.58e-07 \\
    Prod. & Reactions Given & 5.48e-06 & 4.57e-05 & 5.80e-07 \\
    \bottomrule
  \end{tabular}
\end{table*}

\begin{table*}[t]
  \caption{Study 2: generalized eta-squared effect sizes for the two-way mixed ANOVA reported in Table~\ref{tab:s2anova} (60-day window).}
  \label{tab:ges2}
  \small
  \begin{tabular}{llrrr}
    \toprule
    Group & Measure & Cond. & Time & Inter. \\
    \midrule
    Manip. & Offensive Posts Viewed & 0.003 & 0.002 & 0.002 \\
    Manip. & Offensive Comments Viewed & 0.001 & 9.49e-04 & 0.001 \\
    Sessions & Platform Sessions & 4.81e-06 & 2.60e-06 & 2.53e-07 \\
    Consump. & Post Views & 8.13e-06 & 2.77e-08 & 2.86e-09 \\
    Consump. & Comment Views & 1.31e-05 & 7.43e-06 & 4.04e-07 \\
    Consump. & Ad Views & 7.37e-06 & 1.97e-05 & 3.87e-06 \\
    Consump. & Viewed Posts Reported & 3.43e-05 & 1.25e-07 & 3.11e-08 \\
    Consump. & Avg Toxicity of Posts Viewed & 8.77e-09 & 3.29e-05 & 1.58e-06 \\
    Consump. & Avg Toxicity of Comments Viewed & 2.87e-06 & 4.50e-05 & 4.07e-06 \\
    Prod. & Posts Created & 6.10e-06 & 6.05e-05 & 2.19e-06 \\
    Prod. & Comments Created & 4.64e-05 & 5.93e-10 & 8.21e-08 \\
    Prod. & Total Reactions & 2.23e-05 & 3.92e-09 & 4.08e-07 \\
    Prod. & Offensive Posts Created & 3.33e-07 & 3.00e-06 & 3.00e-06 \\
    Prod. & Offensive Comments Created & 2.77e-06 & 3.42e-08 & 4.92e-06 \\
    Prod. & Posts Reported by Others & 1.12e-05 & 1.28e-07 & 8.91e-10 \\
    \bottomrule
  \end{tabular}
\end{table*}

\end{document}